\documentstyle[aps,epsfig]{revtex} 

\newcommand{\tQ}{\tilde{Q}}

\font\zfont=cmss10 
\newcommand{\zz}{\hbox{\zfont Z \kern-.4emZ}} 
\newcommand{\unit}{\hbox{\zfont 1 \kern-.7em1}} 
\newcommand{\step}{\hbox{\zfont S \kern-.75emS}} 
\newcommand{\stepT}{\hbox{\zfont S \kern-.75emS}^T}

\newcommand{\be}{\begin{equation}}
\newcommand{\ee}{\end{equation}}
\newcommand{\ber}{\begin{eqnarray}}
\newcommand{\eer}{\end{eqnarray}}
\newcommand{\ba}{\begin{array}}
\newcommand{\ea}{\end{array}}
\newcommand{\bacc}{\begin{array}{cc}}
\newcommand{\baccc}{\begin{array}{ccc}}
\newcommand{\bacccc}{\begin{array}{cccc}}
\newcommand{\lp}{\left(}
\newcommand{\rp}{\right)}
\newcommand{\ls}{\left[}
\newcommand{\rs}{\right]}
\newcommand{\lb}{\left\{}
\newcommand{\rb}{\right\}}

\newcommand{\tr}{{\mathrm tr}}

\begin{document}

\title{
An improved random matrix model for the chiral phase transition 
in QCD at finite chemical potential
}

\author{M. \'A. Hal\'asz}
\address{Department of Physics and Astronomy, University of Pennsylvania \\
Philadelphia, PA 19104-6396 }

\newcommand{\preprintno}{\normalsize UPR-914-T}

\maketitle

\begin{abstract}
We consider a lattice-inspired random matrix model for the QCD chiral phase
transition at finite chemical potential.
Useful features of the usual RMM for QCD at finite $\mu$ are 
reobtained, some being brought closer to their lattice equivalent.
The simple physical requirement of a vanishing quark number density in
the broken phase is fulfilled in the limit of a large number
of timeslices.

It is argued that the suppression of the partition function at nonzero $\mu$ 
in the broken phase, seen in the usual RMM, is possibly present in 
lattice simulations and is simply a result of the discretization in time.
\end{abstract}


\section{Introduction}

The random matrix model (RMM) for chiral symmetry breaking in QCD at finite chemical
potential has been
introduced by Stephanov \cite{mishaquench}. Significant qualitative insight
 has resulted from
employing the RMM as a schematic model of [lattice] QCD at finite chemical
potential \cite{qcdmu}. 
The failure of the quenched approximation in lattice QCD at finite $\mu$
 was explained in this context by the non-analyticity of the
$N_f \rightarrow 0$ limit \cite{mishaquench}. An extension of the RMM to
include both temperature and chemical potential offers a simple description
of the tricritical point in QCD \cite{tricritical}.

It was noted early on that the partition function of the RMM at finite $\mu$ 
is suppressed in the low density phase \cite{mupaper}, specifically, it
depends on $\mu$ like $e^{-\mu^2 N}$ where $N$ is the size of the  random
matrix. The suppression
was assigned to averaging over the complex phase of the fermion determinant. 
The study of the Glasgow method \cite{glasgow} applied to the same model
\cite{RMglasgow} showed that the method requires exponentially large 
ensembles for convergence. This property was directly traced to the smallness
of the partition function close to the critical chemical potential.

However, the decrease of the partition function with $\mu$ \em below \em
the critical chemical potential in unphysical. One would expect that in
the low density phase there is no $\mu$ dependence so that the number
density $\langle n \rangle = \partial_\mu \ln Z $ is identically zero.
In this paper we propose a different way to introduce the chemical potential
dependence in a RMM, more similar to lattice calculations.

\bigskip

In the next section we introduce the model and work out the
effective theory. The model copies the time-structure of the lattice
Dirac operator at finite chemical potential. Using standard RMT
methods and some algebraic manipulations we reduce the initial
integral over an $N \times N$ matrix to one over an $N_f \times N_f$ matrix,
where $N_f$ is the number of quark flavors.
Sec.III is devoted to the analysis of the partition function for one flavor, 
using the saddle point approximation. We identify the phase structure and
show the central result of this paper, the $\mu$-independence of the
low density phase in the limit of a large number of timeslices. 
In Sec.IV we show numerical results on the eigenvalue distribution
of the model Dirac operator and the zeros of the partition function.
The eigenvalues show a strong qualitative similarity to actual lattice results.
The zeros of the partition function trace out the boundary between the 
two phases in the complex plane of the parameters, reinforcing our 
saddle point analysis.
We summarize and discuss our findings in the final section. The Appendix 
contains technical details used in the derivation of the effective partition
function.

\section{Model}

\subsection{Definition}

Consider a partition function of the generic form
\ber
Z_N(m,\mu)~=~\int {\cal D} C e^{- N \tr C C^\dagger}
\det \lp 
\bacc 
   m \otimes \unit_N                & i g C + W(\mu) \\ 
   i g C^\dagger - W^\dagger (-\mu) & m \otimes \unit_N 
\ea 
\rp^{N_f}~~.
\eer
which is dictated by the chiral structure, mass ($m$) and chemical potential ($\mu$)
dependence of the QCD partition function. 
The integration is over an $N \times N$ complex matrix $C$, where ${\cal D} C$ is
the Haar measure.
The size parameter $g$ can
be seen as a measure of the strength of the interaction. 

The usual choice for the $\mu$-dependence, $W(\mu)=\mu$ is also the  simplest one 
\cite{mishaquench,tricritical,mupaper}.
We wish to build a model whose chemical potential dependence mimics
more closely that of the lattice, where the chemical potential enters through factors of
$e^{\pm \mu}$ multiplying the forward and backward links \cite{qcdmu}. 
Therefore we imagine that we have
$N_t$ timeslices with $N_s$ points in each of them, so that $N = N_t N_s$. In the absence of 
a temperature dependence, we choose
\ber
W(\mu)~=~  \lp e^\mu \step_{N_t} \otimes \unit_{N_s} - 
                  e^{-\mu} \stepT_{N_t} \otimes \unit_{N_s} \rp~~,
\eer
where $\step_N$ is the forward step matrix of size $N$.
This way each block of size $N_s$ is coupled to the next and to the previous block through
a factor $e^{\pm \mu}$, respectively.

\subsection{Reducing the partition function}

We start by performing a standard Hubbard-Stratonowitz transformation.
This leads to replacing the integration over the complex $N \times N$ matrices $C$ 
with integration over an $N_f \times N_f $ complex matrix $\sigma$.
The size of the matrix under the determinant 
can be reduced by 2 via the identity
$\det \lp \bacc A & B \\ C & D \ea \rp = \det( A D - B D^{-1} C D )$ and taking into account
that the resulting upper and lower right blocks commute, we obtain
\ber
Z~\sim~ \int {\cal D} \sigma e^{- N \tr \sigma \sigma^\dagger}
\det \ls \lp {\cal M} + g \sigma \rp 
         \lp {\cal M} + g \sigma \rp^\dagger 
         \otimes \unit_{N_t} -
         \unit_{N_f} \otimes
          \lp e^{\mu} \step_{N_t} -  e^{-\mu} \stepT_{N_t} \rp^2 
\rs^{N_s}~~.
\label{pfgen}
\eer
Note that the two remaining terms under the determinant may be diagonalized simultaneously.
In the $N_f > 1$ case, one may calculate the exact partition function by switching to
the distribution of the eigenvalues of $\sigma$. In the following we will simply assume 
equal masses and $\sigma$ diagonal, effectively leaving us with one flavor. 
We also assume that $N_t$ is even, $N_t= 2 N_t'$, so the square of the step matrix breaks up 
into $2 \times 2$ blocks, $\step_{N_t}^2 = \step_{N_t'} \otimes \unit_2$, and so we can 
again reduce the size of the matrices by two. 
Our partition function is now
\ber
\label{pfdoubleint}
Z~\sim~ \int d^2 \sigma \sigma e^{- N \sigma \sigma^*}
\det \lb \ls ( m + g \sigma )( m  + g \sigma^* ) +2 \rs  \otimes \unit_{N_t'} 
         - e^{2 \mu} \step_{N_t'} -  e^{-2 \mu} \stepT_{N_t'} 
     \rb^{2 N_s}~~.
\label{pfnf1}
\eer
The matrix under the determinant is of the form considered in the Appendix,
\ber
Q_L(a,x)~=~\lp \bacccc a    &    x   & \cdots & x^{-1} \\ 
                     x^{-1} & \ddots & \ddots & \vdots \\
                     \vdots & \ddots & \ddots &  x \\
                        x   & \cdots & x^{-1} &  a \ea \rp~~.
\eer
with $L=N_t'$, $a=(m+g \sigma)(m+g \sigma^*)+2$, and $x=- e^{2 \mu}$. 
The determinant is then
\ber
\det ( Q ) ~=~ \lambda_1(a)^L + \lambda_2(a)^L - (-)^L  (x^L + x^{-L})~~.
\eer
where $\lambda_{12}(a)$ are the two solutions of $\lambda^2 - a \lambda + 1 = 0$.
For simplicity we will assume that $N_t'$ is even.
By convention, if they are real, we define $\lambda=\lambda_1$ to be the one
larger than $1$ in absolute value.
With all this, we have
\ber
\label{pfeff}
Z~\sim~\int d^2 \sigma e^{-N \sigma \sigma^*} 
\prod\limits_{k=1}^{N_{sec}} \lp
\lambda(a)^{N_t'} + \lambda(a)^{- N_t'} 
- e^{2 \mu N_t'} - e^{- 2 \mu N_t'} \rp^{2 N_s N_f}~~,
\eer
where $N= N_t N_s$, $N_t = 2 N_t'$, $a=(m+g \sigma)(m+g \sigma)+2$.

\section{Phase structure}

Let us consider the case of one flavor, $N_f=1$.
The partition function depends on two thermodynamic 
parameters, the quark mass $m$ and the quark chemical potential, $\mu$.
The derivatives with respect to these two quantities are proportional to the chiral 
condensate $\chi~=~\frac{1}{N} \partial \ln Z_N ( m , \mu ) / \partial m$ and
the quark number density $n ~=~\frac{1}{N} \partial \ln Z_N ( m , \mu ) / \partial \mu $. 
By changing variables to $\tau=\sigma+m$ it is easy to see that 
$\chi = \langle 2 \Re (\sigma)\rangle $. 
For positive $m$,  the integral over the phase of $\tau$ has a saddle point at 
$\arg ( \tau) = 0$, so the integral over $d^2 \tau$ turns into a radial
one. With a further change of variables to $u=\tau^2$, we obtain
\ber
\label{pfumg}
Z(m,\mu,g) \sim e^{-2 N_t' N_s \frac{m^2}{g^2}} 
\int\limits_0^\infty du \lb e^{- N_t' (u - 2 \frac{m}{g} u^{1/2} )} 
Q_{N_t'}(g^2 u +2;e^{-2 \mu} ) \rb^{2 N_s}  .
\eer

\bigskip

Let us first focus on the $m \rightarrow 0$ behavior, by setting 
$m=0$ in (\ref{pfumg}) and bearing in mind that $\chi = 2 \langle \sqrt{u} \rangle$.
We may write the partition function as
\ber
\label{pfsimple}
Z(\mu)~\sim~ \int\limits_{0}^{\infty} d u e^{-N u } 
\lp
\lambda(a)^{N_t'} + \lambda(a)^{- N_t'} 
- 2 \cosh \lp 2 \mu N_t' \rp \rp^{2 N_s}~~,
\eer
where now $a=u g^2+2$ and $N=2 N_s N_t'$.
What is new about our partition function in comparison with the usual RMM 
is that here we have \em two \em
potentially large numbers, $N_t'$ and $N_s'$. 
Since $N_s$ is a large number (in lattice simulations it is of 
the order of $N_t^3$, e.g., $500$ for $N_t=8$), it is justified to
 approximate $Z$ by the saddle point result corresponding to the absolute
maximum of the function $\Phi$,
\ber
\lb e^{- N_t' u} \ls \lambda(u g^2+2)^{N_t'} + \lambda(u g^2+2)^{-N_t'}
 -2 \cosh(2 \mu N_t') \rp \rb^{2 N_s}
= \Phi(u,g,N_t',\mu)^{2 N_s}~~.
\eer
In actual lattice simulations the number
of timeslices $N_t=2 N_t'$, is not very large, for instance, in \cite{glasgow}, $N_t=2 N_t'=8$.

For small $N_t$ values out model can be worked out in the traditional manner.
The integrand for $N_t'=2$ and $g=1$ is rewritten as 
\ber
\lb e^{- 4 u} \lp \lp u+2 \rp^2 - 4 \cosh(2 \mu )^2 \rp \rb^{2 N_s}
= \Phi(u,g,2,\mu)^{2 N_s}~~.
\eer
 The stationary points of the function
are at $ \bar{u}_{12}= \lp - 3 \pm \sqrt { 9 + 8 \cosh (4 \mu) } \rp /2$. 
Since the integration is from 
$u=0$, the negative stationary point does not matter, but one needs to take into
consideration the endpoint, $u=0$. Altogether, we find two phases, one with 
$\chi=\langle \sqrt{u} \rangle=0$ for large
$\mu$ and one with $\chi \neq 0$ for small $\mu$. They are separated by a first order 
phase transition at $\mu_c=0.1966$, when 
$\Phi(0,g=1,N_t'=2,\mu_c)=\Phi(\bar{u}_1,g=1,N_t'=2,\mu_c)$.
Similarly to the usual random matrix model at finite $\mu$, 
the quark density starts off
from zero at $\mu=0$, decreases until $\mu_c$, where it has a finite upward 
jump, and then continues to decrease. 

For $N_t'$ larger than $4$ a closed form solution is not possible. 
However, the general
features of the function $\Phi$ for $u \geq 0$ are similar, in the sense that 
$\Phi(u,g,N_t',\mu=0)$
starts off at zero, increases and has exactly 
one maximum, after which it asymptotically 
decreases again to zero. For $\mu > 0$, the value at 
$u=0$ is negative, but the initial 
increase and the unique maximum are preserved for moderate $\mu$ values.

\begin{figure}
\centerline{
\vbox{
\psfig{file=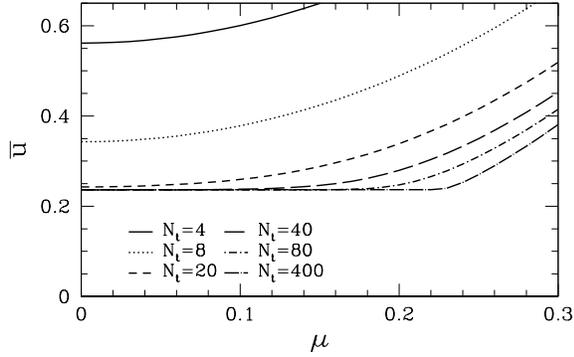,width=8cm,angle=0}
\vspace{-2.5cm}}}
\caption{The location $\bar{u}$ of the maximum of $\Phi(u,g=1,N_t',\mu)$ with respect to $u$
as a function of $\mu$ for various numbers of timeslices, $N_t=2 N_t'$.
As $N_t$ increases, the $\mu$-dependence becomes negligible, leading to a low density
phase with identically zero quark number.
\label{umax}
}
\end{figure}

\begin{figure}
\centerline{
\vbox{
\psfig{file=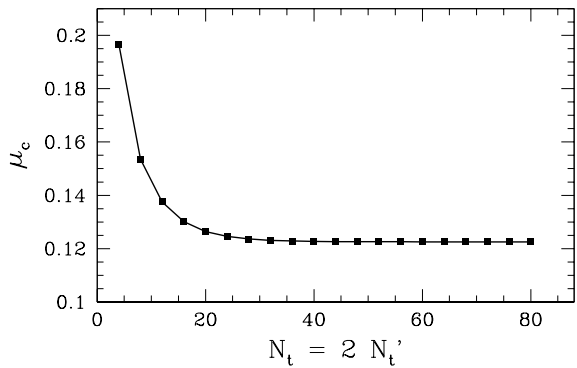,width=8cm,angle=0}
\vspace{-2.5cm}}}
\caption{
The value of $\mu_c$ for various numbers of timesteps $N_t=2 N_t'$, at $g=1$, $m=0$. 
We define $\mu_c$ as the value of mu at which the magnitude
of $\Phi$ at $u=0$ equals the value of $\Phi$ at its maximum.
\label{muvsn}
}
\end{figure}

We plot $\bar{u}$, the location $\bar{u}$ of the maximum of $\Phi(u)$,
as a function of $\mu$ for various $N_t$, in Figure \ref{umax}.
We can see how this quantity, which is directly related to
the chiral condensate in the broken phase, $\chi = 2 \sqrt{\bar{u}}$, becomes
increasingly independent of $\mu$ as $N_t$ increases.  
In figure \ref{muvsn} we plot the critical chemical potential $\mu_c$ for various 
values of $N_t = 2 N_t'$. Again, $\mu_c(N_t)$ quickly converge towards a limiting 
value.


\bigskip

Assuming $N_t$ large, things become very clear. The quantities $\lambda(u g^2+2)^{\pm 1}$,
solutions of a quadratic equation,
 are always real (for positive $u$), and  one of them, $\lambda = \lambda_1$ is
larger than $1$, and increases with $u$. Therefore the function 
$e^{- N_t' u} \lambda(u g^2+2)^{N_t'}$ has a unique maximum (for positive $u$) at 
$\bar{u}=\sqrt{1+ 4 / g^4}-2/g^2$  ($\bar{u}=0.236\ldots$ for $g=1$), 
corresponding to $\bar{\lambda}=g^2/2 + \sqrt{1+g^4/4}$ 
($\bar{\lambda}=1.618\ldots$ for $g=1$). 
For a significantly large $N_t'$ this maximum is not disturbed by  the $1/\lambda$ term. 
In order for the $\cosh( 2 \mu N_t')$ term to matter, $\exp ( 2 \mu )$ has to be comparable to
$\bar{\lambda}$, which corresponds to $\mu$ around $\mu_1 = 0.2406\ldots$. 
However, this does not come into play since
the maximum at $\bar{u}$ competes with the absolute value of $\Phi$ at $u=0$, and
the approximate condition for the equality of the $\Phi$ values is
\ber
exp(- 2 \mu_c) = exp( - \bar{u} ) \bar{\lambda}~~,
\eer
which gives $\mu_c=0.1226\ldots < \mu_1$ . Therefore the maximum at $\bar{u}$ ceases to
dominate before its location is significantly influenced by $\mu$.
The transition between the maximum at $\bar{u}$ and the value at $u=0$ becomes sharper
as $N_t$ increases. 

\begin{figure}
\centerline{
\vbox{
\psfig{file=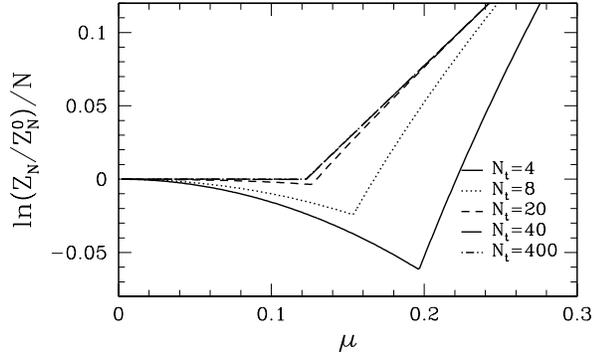,width=8cm,angle=0}
\vspace{-2.5cm}}}
\caption{
The absolute value of the logarithm of partition function at $m=0$, 
normalized to its value $Z_N^0 = Z_N(\mu=0)$ and divided by 
$ N=N_s N_t = 2 N_s N_t' $, in the saddle point approximation with respect to $N_s$,
for various values of $N_t'=N_t/2$. 
The value $N_t=2 N_t'=8 $ corresponds to usual lattice simulations.
Notice that for larger $N_t$ values the partition function becomes practically
flat below $\mu_c$.\label{lpfunc}
}
\end{figure}

\begin{figure}
\centerline{
\vbox{
\psfig{file=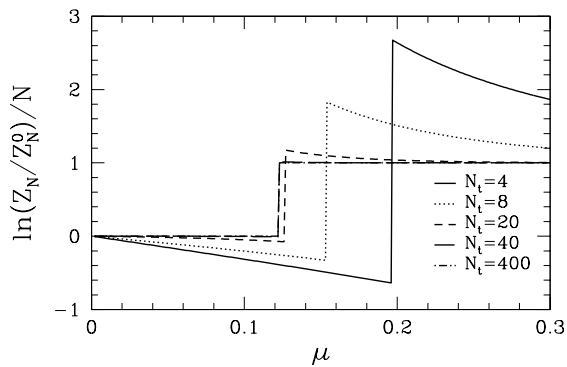,width=8cm,angle=0}
\vspace{-2.5cm}}}
\caption{
The value of the number density versus the chemical potential, at $m=0$,
corresponding to the partition function plotted in Fig.\ref{lpfunc}.
\label{numdens}
}
\end{figure}

\begin{figure}
\centerline{
\vbox{
\psfig{file=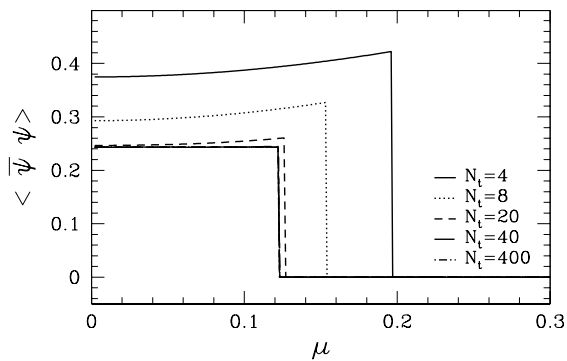,width=8cm,angle=0}
\vspace{-2.5cm}}}
\caption{
The value of the chiral condensate versus the chemical potential, at $m=0$,
corresponding to the partition function plotted in Fig.\ref{lpfunc}.
The number density and the chiral condensate mutually exclude each other.
\label{cond}
}
\end{figure}

In Figures \ref{lpfunc},\ref{numdens} and \ref{cond} we plot the saddle
 point result 
for the partition function and the two main observables, the number density
and the chiral condensate, for a large number of spatial points ($N_s >> 1$), 
and for various $N_t'$ values.
In Figure \ref{lpfunc} we plot the logarithm of the absolute value of the partition
function, normalized to its value at $\mu=0$, and divided by $N=N_s N_t=2 N_s N_t'$.
As we have discussed, the $\mu$-dependence for small $N_t$ values is very similar to
that seen in the usual random matrix model.
The partition function decreases in magnitude with increasing $\mu$ up to the 
critical chemical potential. As a matter of fact, this suppression persists for all
finite $N_t$ but it drops quickly in magnitude as $N_t$ increases, so that for $N_t=40$
the partition function is practically flat between $0$ and $\mu_c$.
The same trend can be seen in the plot of the number density \ref{numdens}. The
decrease of the partition function is translated into a negative number density,
which is certainly unphysical. This feature was present in the usual random matrix models,
and was discarded as an artifact of the model. Finally, the chiral condensate is also
sensitive to $\mu$ even in the low density (or broken) phase. This corresponds directly 
to the fact that the maximum of $\Phi$ moves as $\mu$ varies.

An important conclusion we may draw from our analysis is that 
the physical requirement of $\mu$-independence of the low density phase is ensured by 
a large $N_t$, i.e., a fine enough time discretization.
Another interesting point is that the suppression is always present to
some extent. This is probably so for Glasgow simulations at finite $\mu$,
where the number of timeslices used was as low as $N_t=8$. But even for higher
numbers of timeslices, the suppression is enhanced by a factor of $N_t N_s$.
Although the suppression decreases significantly faster than $1/N_t$, the
factor $N_s$ is typically very large in lattice simulations, since it is of the
order $N_t^3$. 
Therefore it is not unlikely that the unphysical artifact seen in the usual RMM,
leading to a negative number density, is also present in traditional lattice 
simulations. In our analysis of the Glasgow method, we have found a connection between
the failure of the Glasgow method and the suppression of the partition function.

\bigskip

\begin{figure}
\centerline{
\vbox{
\psfig{file=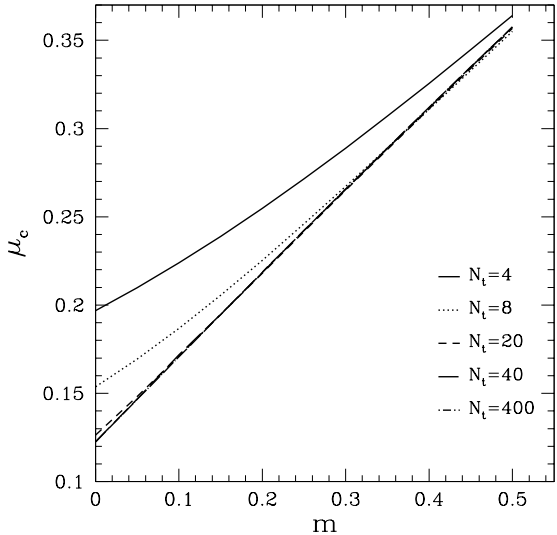,width=8cm,angle=0}
\vspace{0cm}}}
\caption{
The value of the critical chemical potential $\mu_c$ as a function of the
quark mass $m$, for various $N$ values. The critical chemical potential
increases with the mass. 
\label{muvsm}
}
\end{figure}

For $m\neq 0$, $g \neq 1$ the analysis is similar. The $N_t \rightarrow \infty$ limit
leads to a sharper transition and increasingly $\mu$-independent broken phase.
The mass dependence of the partition function is similar to that in the old model. 
For a positive mass, the exponential in $\Phi$ has a maximum at $u_0=m^2/g^2$, which
competes now with the maximum at $\bar{u}$. The location of $\bar{u}$ is also shifted
upwards and so is the size of the maximum, while the magnitude of $\Phi(u_0)$ is 
only marginally influenced by the mass. As a result, there is a first order phase
transition at any $m$, and $\mu_c$ increases with $m$. 
The critical curve in the $m - \mu$ plane is shown in 
Figure \ref{muvsm}, for various $N_t'$. 

\begin{figure}
\centerline{
\vbox{
\psfig{file=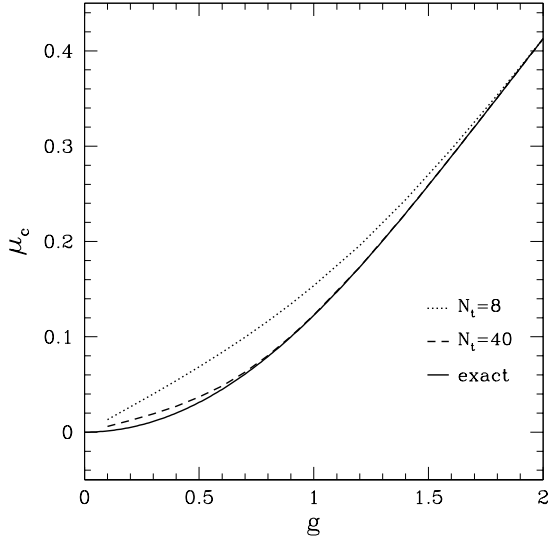,width=8cm,angle=0}
\vspace{0cm}}}
\caption{
The value of the critical chemical potential $\mu_c$ as a function of the
parameter $g$, for various $N$ values and in the limiting case. 
The critical chemical potential
increases with $g$ and tends to zero when $g$ vanishes. 
\label{muvsg}
}
\end{figure}

The dependence on the size parameter $g$ is also easy to understand. 
For small $g$ values,
$\mu_c$ approaches zero, since the value of the maximum of $\Phi$ approaches $1$.
For large $g$, $\bar{u}$ approaches $1$ and the value of the maximum grows
like $g^2$, so $\mu_c(g)$ increases logarithmically. 
In fact, for zero mass we can compute the limiting curve $\mu_c(g)$ at $N_t \rightarrow \infty$,
in a closed form: $\mu_c= \frac12 ( \ln \bar{\lambda} - \bar{u} )$ where 
$\bar{\lambda} = \frac{g^2}{2} + \sqrt{1 + \frac{g^4}{4}}$ and 
$\bar{u}= -\frac{2}{g^2} + \sqrt{1 + \frac{4}{g^4}}$. This curve and $\mu_c(g)$ for $N_t=8$
and $N_t=40$ are plotted in Figure \ref{muvsg}.

The $m$ and $g$-dependence of 
the phase boundaries is nicely illustrated in the next section, by the partition
function zeros. In particular, for small $g$ the zeros are sitting on a contour
around the roots of unity.
The contours plotted in Figure \ref{zeros} are obtained by comparing the value of
$\Phi$ at $u=0$ and some complex $\mu$ with that of $\Phi(\bar{u},\mu)$, but with 
$\bar{u}$ obtained at $\Re{\mu=0}$. Even so, the agreement with the line of partition
function zeros is very good.

\section{Numerical results}

\subsection{Eigenvalue distribution}

It is widely accepted now that the distribution of the eigenvalues in this
class of non-hermitian random matrix models is connected to the phase structure 
defined by a partition function where the determinant has been replaced by its 
absolute value \cite{mishaquench}. While it is likely that the model under consideration
is no exception, we will only present numerical results here regarding the distribution
of the eigenvalues.
These are qualitatively similar to lattice eigenvalue distributions seen in the
literature \cite{glasgow,latteval}. 

\begin{figure}
\centerline{
\hbox{
\psfig{file=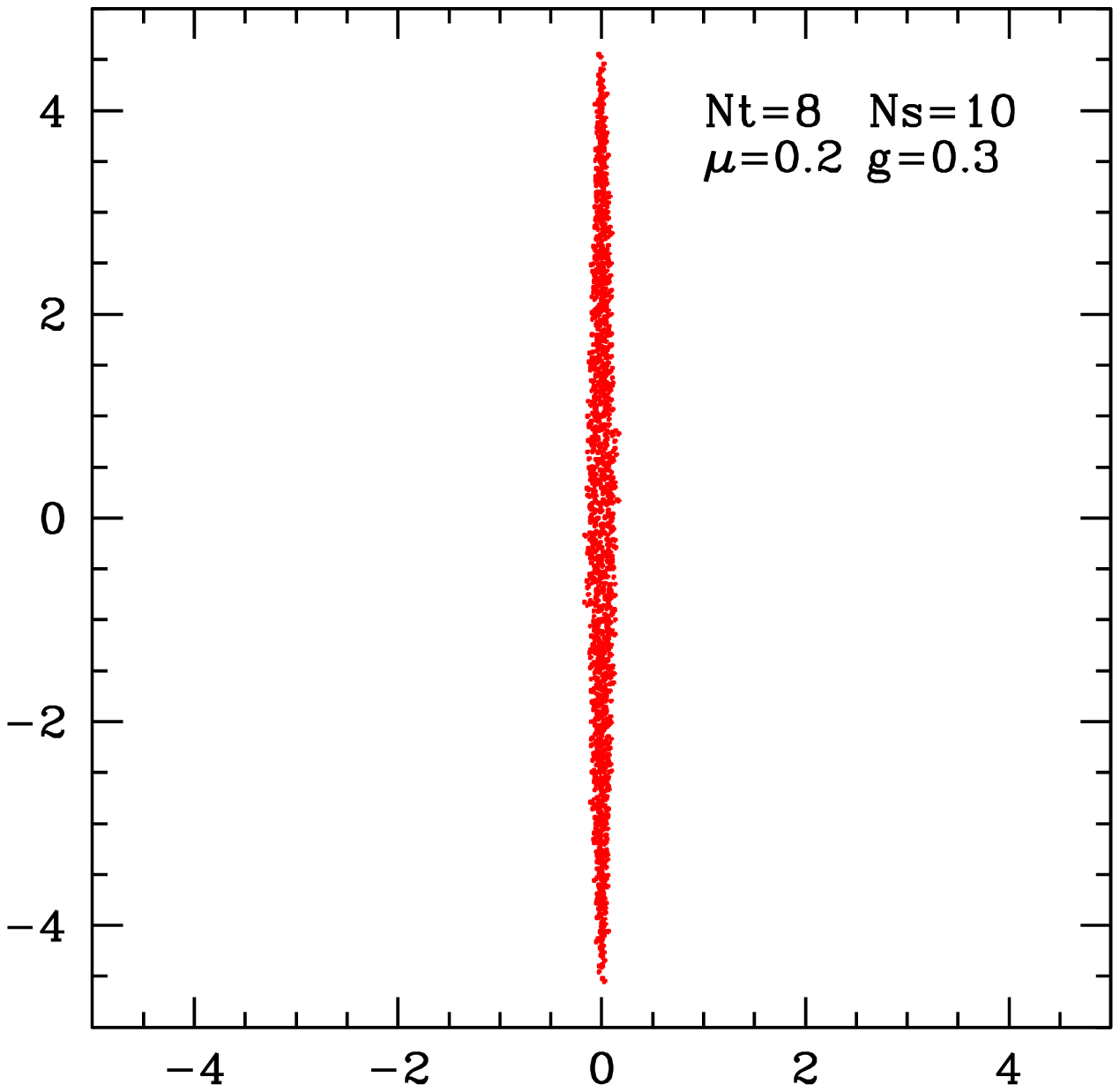,width=8cm,angle=0}
\hspace{-3.0cm}
\psfig{file=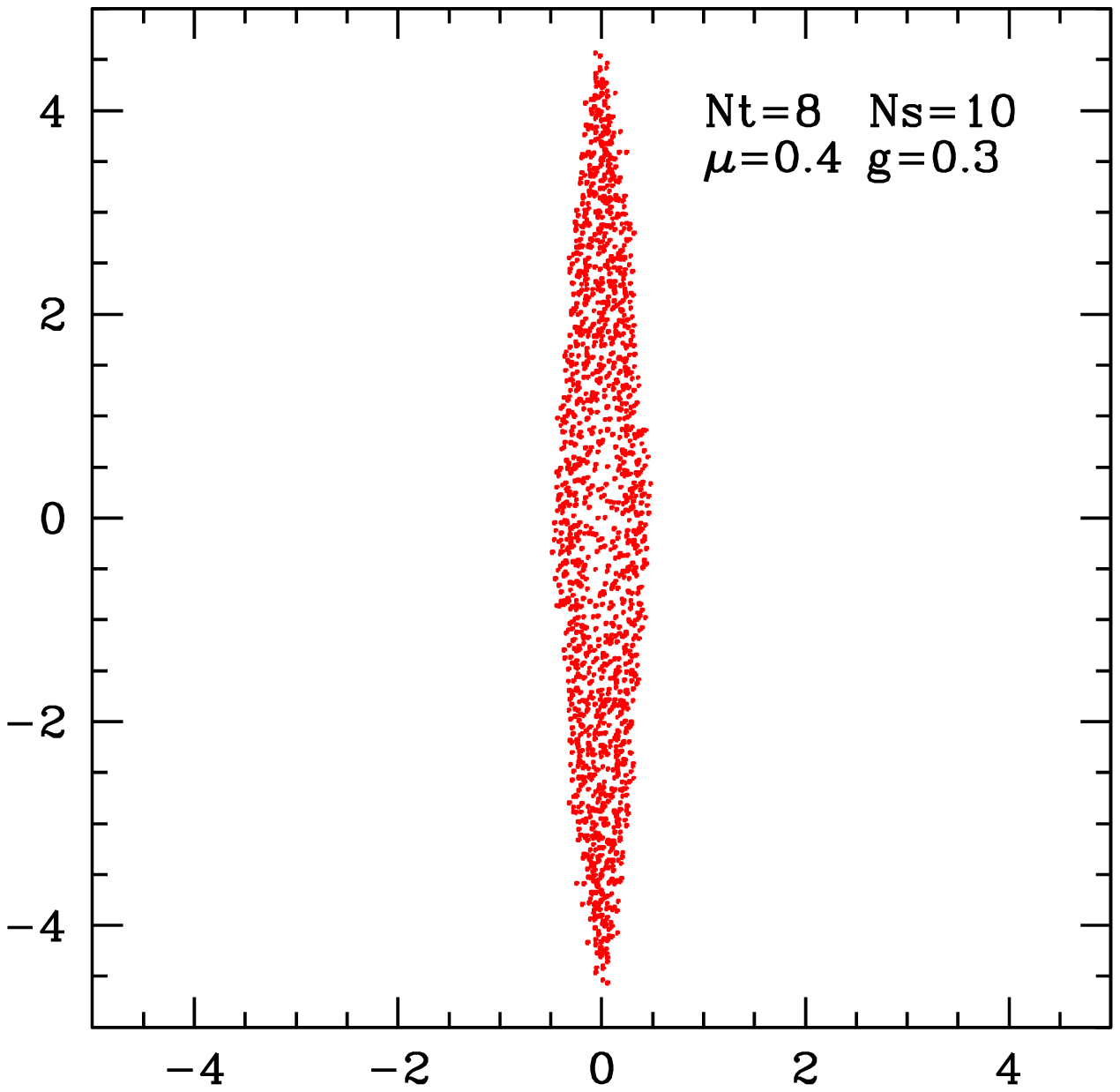,width=8cm,angle=0}
\hspace{-3.0cm}
\psfig{file=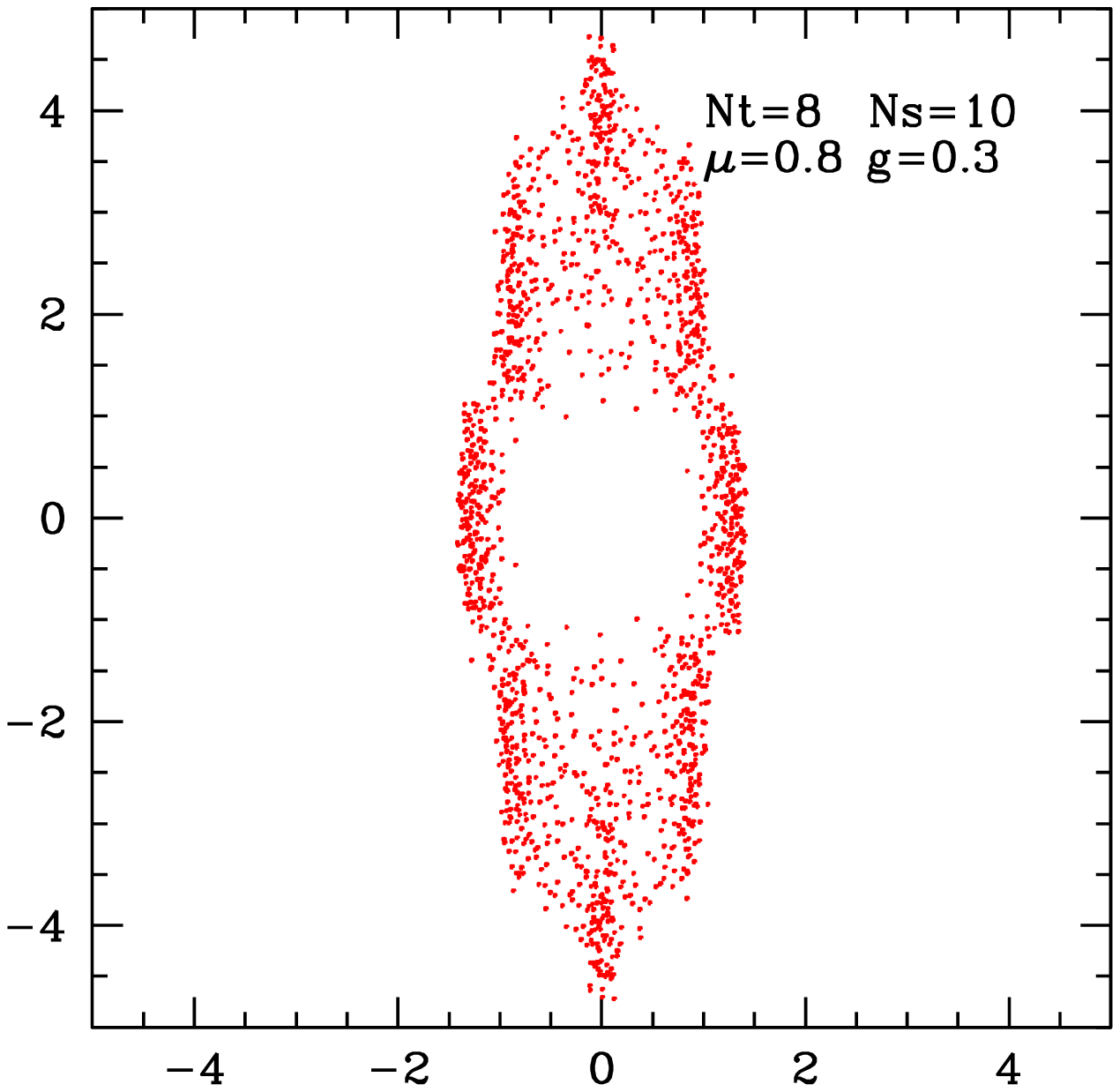,width=8cm,angle=0}
}
\vspace{-2.0cm}
}
\caption{
Eigenvalues in the mass plane for $N_t=8$, $N_s=10$, $g=0.3$, 
$\mu=0.20$, $0.40$, and $0.80$ respectively, from left to right.
The initial line of eigenvalues along the imaginary axis widens
in the real direction. For combinations of large $\mu$ and small
$g$ the eigenvalues are grouped around $N_t$ points located symmetrically
inside the ellipsoidal cloud. 
\label{evalsM_mu}
}
\end{figure}

The eigenvalues in the complex $m$ plane are straightforward to define. They are 
the eigenvalues of the model Dirac matrix
\ber
D~=~\lp \bacc 0 & i C + e^\mu \step_{N_t} \otimes \unit_{N_s} 
                   - e^{-\mu} \step_{N_t}^T \otimes \unit_{N_s} \\
      i C^\dagger +e^\mu \step_{N_t} \otimes \unit_{N_s} 
                   - e^{-\mu} \step_{N_t}^T \otimes \unit_{N_s} & 0 \ea \rp~~.
\eer
Because of the chiral structure, they must come in $\pm$ pairs.
In Figure \ref{evalsM_mu} we plot three sets of eigenvalue corresponding to ensembles of $10$ matrices
with $N_t=8$ and $N_s=10$ for a relatively small value of $g=0.3$ and various nonzero $\mu$ values.
We recognize the generic pattern with the eigenvalues originally on the imaginary axis (for $\mu=0$),
which then spread out gradually in the complex plane as $\mu$ increases. The qualitative 
similarity with lattice eigenvalues is stronger than for the usual RMM, since here the cloud remains
contiguous and only the center is depleted of eigenvalues for large $\mu$, similarly to \cite{latteval}, while in the usual RMM the cloud eventually breaks up into two distinct pieces which are pushed
away along the real axis \cite{mupaper}.

\begin{figure}
\centerline{
\hbox{
\psfig{file=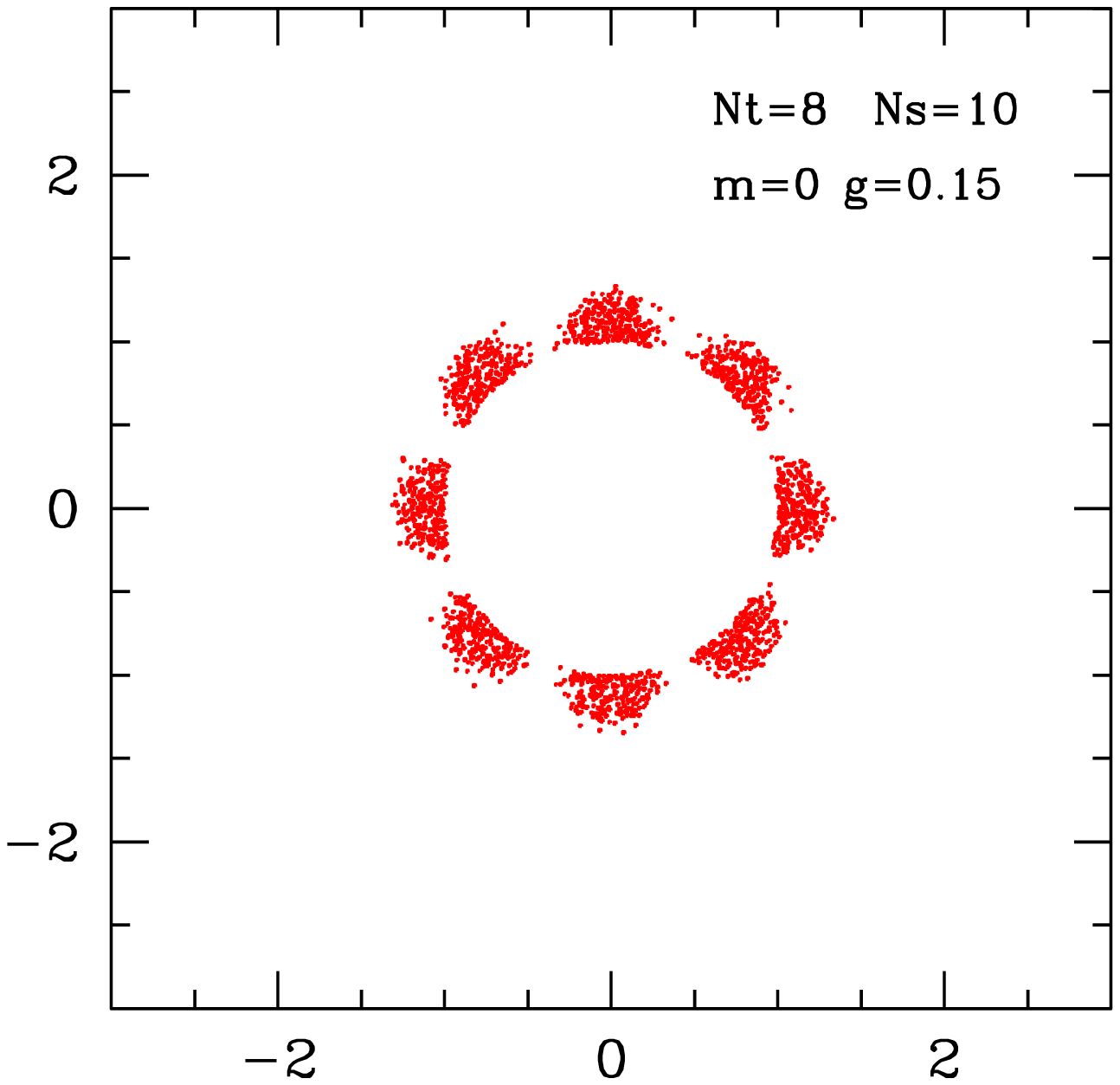,width=8cm,angle=0}
\hspace{-3.0cm}
\psfig{file=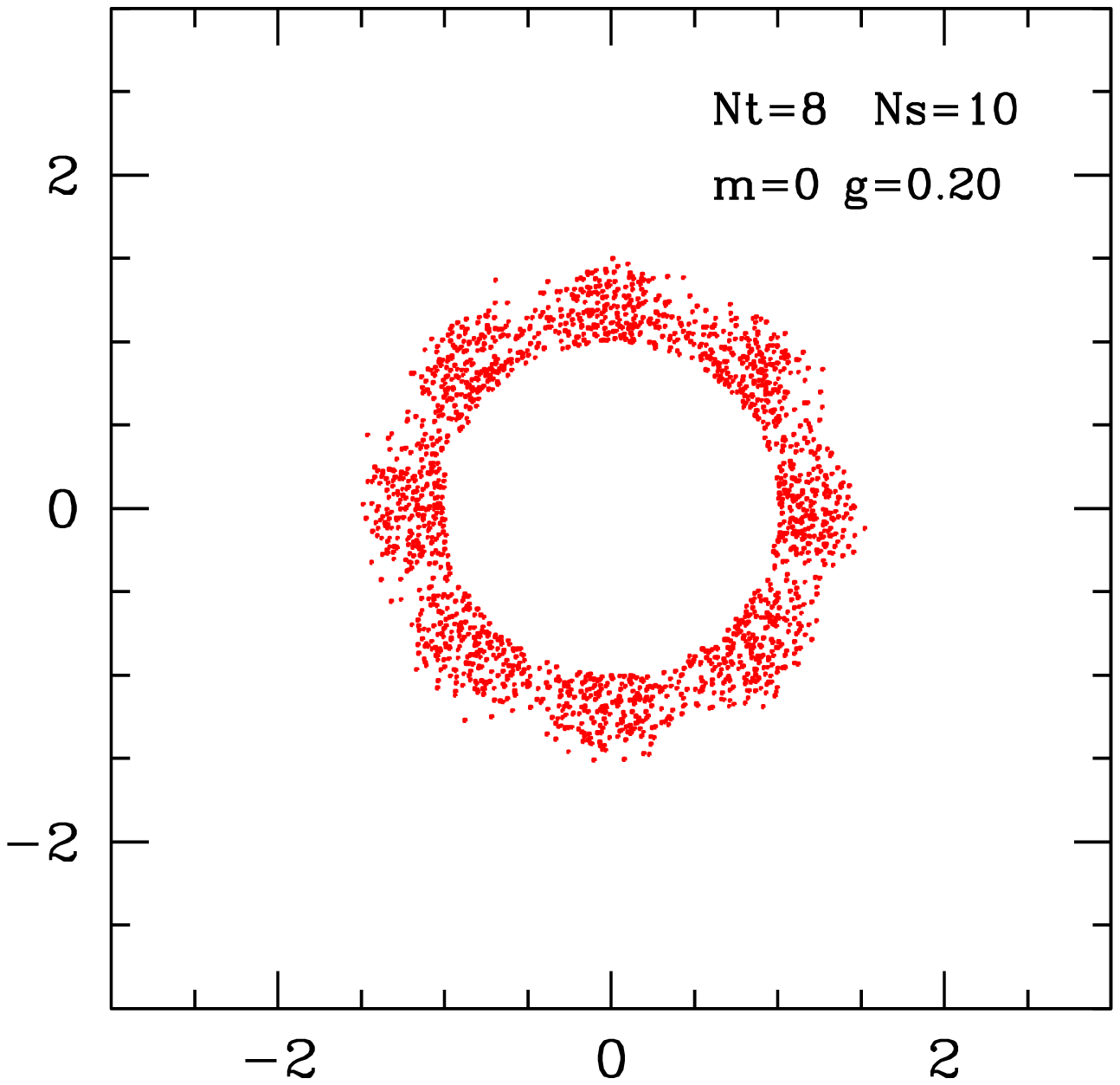,width=8cm,angle=0}
\hspace{-3.0cm}
\psfig{file=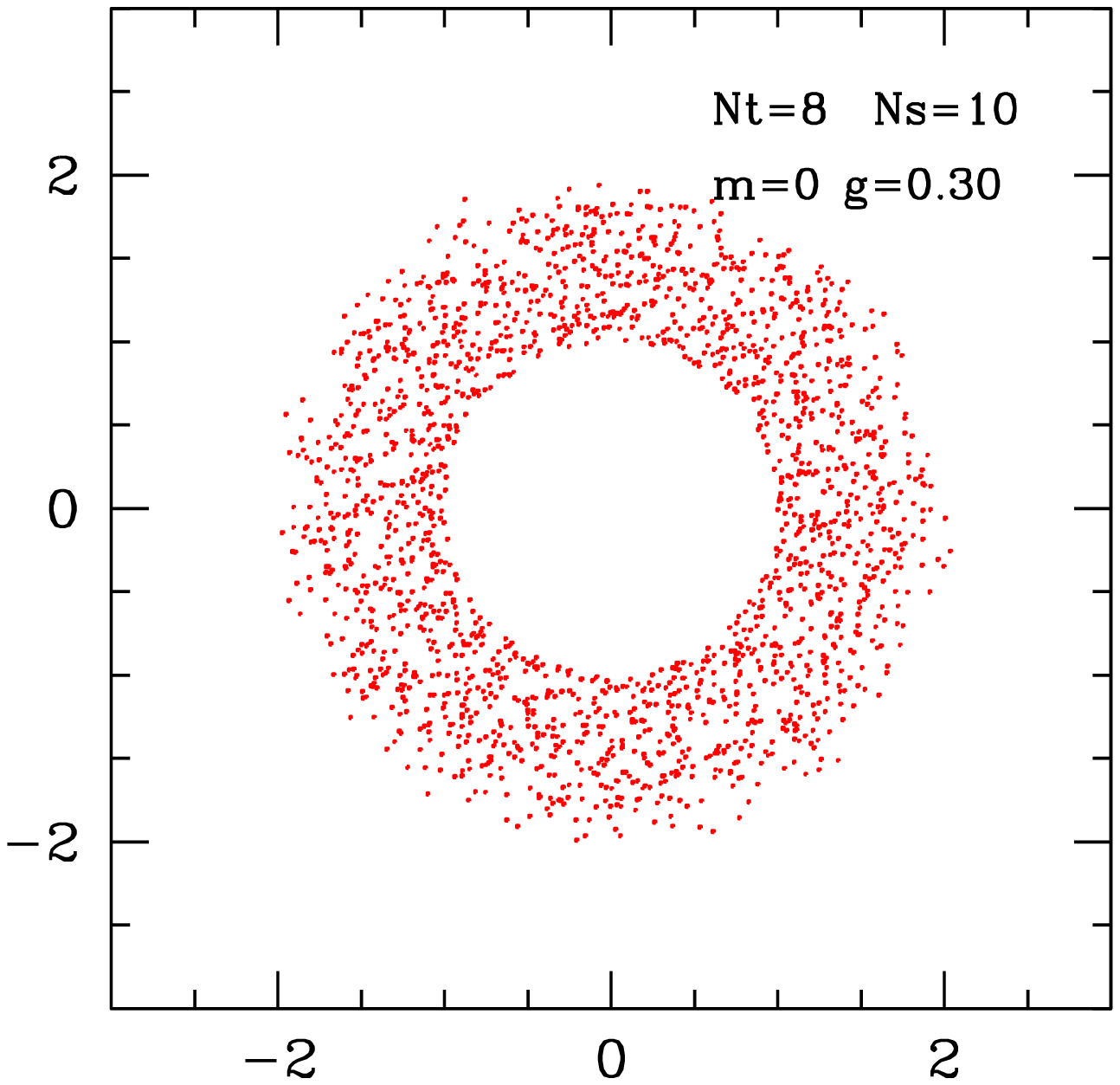,width=8cm,angle=0}
}
\vspace{-2.0cm}}
\caption{
Eigenvalues in $\xi_k$ the fugacity plane for $N_t=8$, $N_s=10$, $m=0$ and 
$g=0.15$, $0.20$, and $0.30$ respectively, from left to right.
Only the ones with $| \xi_k | \leq 1$ are plotted.
For small $g$ the eigenvalues are concentrated around the roots of unity
and merge into a circular band as $g$ increases.
\label{evals_g}
}
\end{figure}

The eigenvalues corresponding to the $\mu$ dependence are perhaps 
more interesting from a practical
point of view. The natural variable is the fugacity $\xi=e^\mu$. We are interested in
eigenvalues in the complex fugacity plane, corresponding more closely to the ones calculated
in the Glasgow method \cite{glasgow} in lattice simulations. We seek the complex values 
of the fugacity which cancel the fermion determinant $D+m$ for a given random matrix
'configuration' $C$. We apply Gibbs' trick \cite{gibbs} to construct a matrix whose
eigenvalues are explicitly the zeros in the fugacity plane. Let us define the matrices
\ber
G~&=&~ \lp \bacc m & i C \\ i C^\dagger & m \ea \rp \nonumber\\
V ~&=&~ \lp \bacc 0 & \step_{N_t} \\ \step_{N_t} & 0 \ea \rp \nonumber\\
\bar{V} ~&=&~ \lp \bacc 0 & \step_{N_t}^T \\ \step_{N_t}^T & 0 \ea \rp ~~.
\eer
so that the fermion matrix is written $D+m = G + \xi V - \xi^-1 \bar{V} $.
It is straightforward to check that
\ber
\det (D + m ) ~=~ \det \lp \xi^{-1} V \rp
\det \lp \bacc G \bar{V} + \xi \otimes \unit_N  & \bar{V} \\
     \bar{V} & \xi \otimes \unit_N \ea \rp~~.
\eer
where we again used the formula for reducing the size of the determinant by $2$.
Hence we are interested in the eigenvalues of the propagator matrix
\ber
F(m)~=~\lp \bacc G \bar{V} & \bar{V} \\ \bar{V} & 0 \ea \rp~=~
\lp \bacc G  & \unit_{N_t} \\ \unit_{N_t} & 0 \ea \rp
\lp \bacc \bar{V}  & 0 \\ 0  & \bar{V} \ea \rp
.
\eer
If the size $g$ of the random matrix is small, the $\xi$-eigenvalues 
are just the $N_t$ order roots of unity. As $g$ increases, they are located in a 
cloud around those values. The size of the cloud increases with $g$, to the point
where the eigenvalues merge into a ring around zero. In figure \ref{evals_g}
we plot only the eigenvalues larger than $1$ in absolute value (since if $\xi$
is an eigenvalue, so is $1/\xi$).

\begin{figure}
\centerline{
\hbox{
\psfig{file=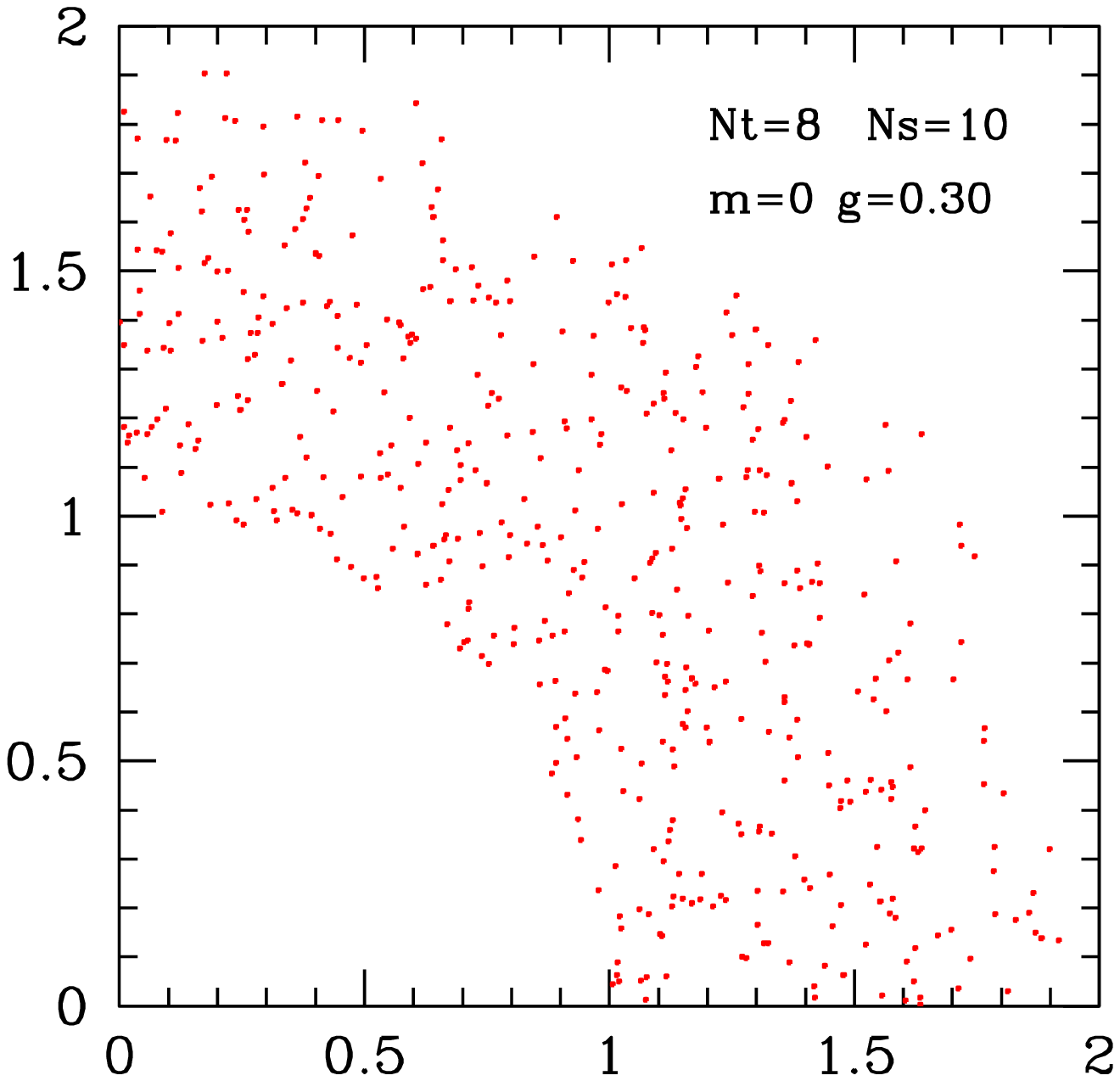,width=8cm,angle=0}
\hspace{-3.0cm}
\psfig{file=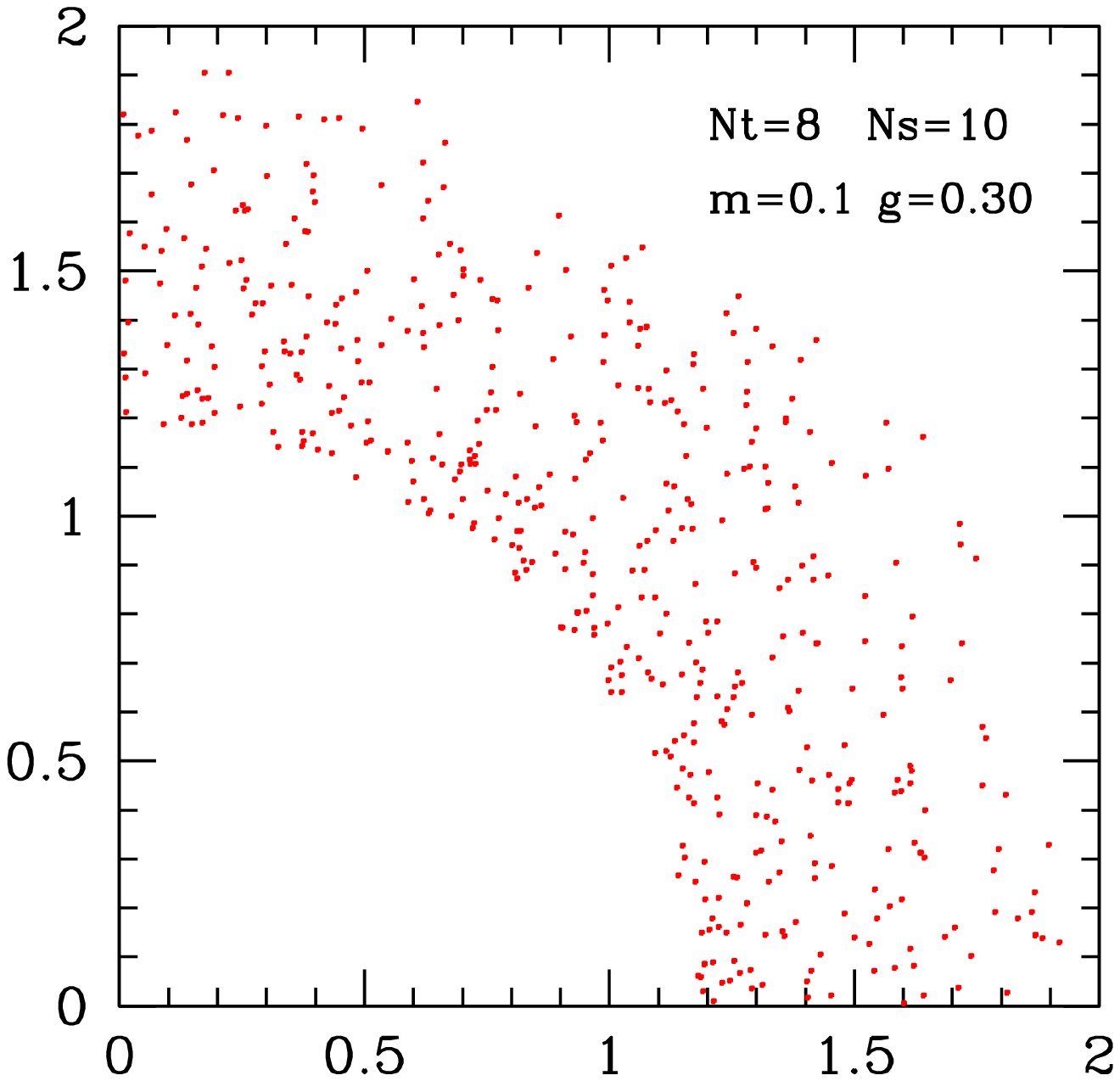,width=8cm,angle=0}
\hspace{-3.0cm}
\psfig{file=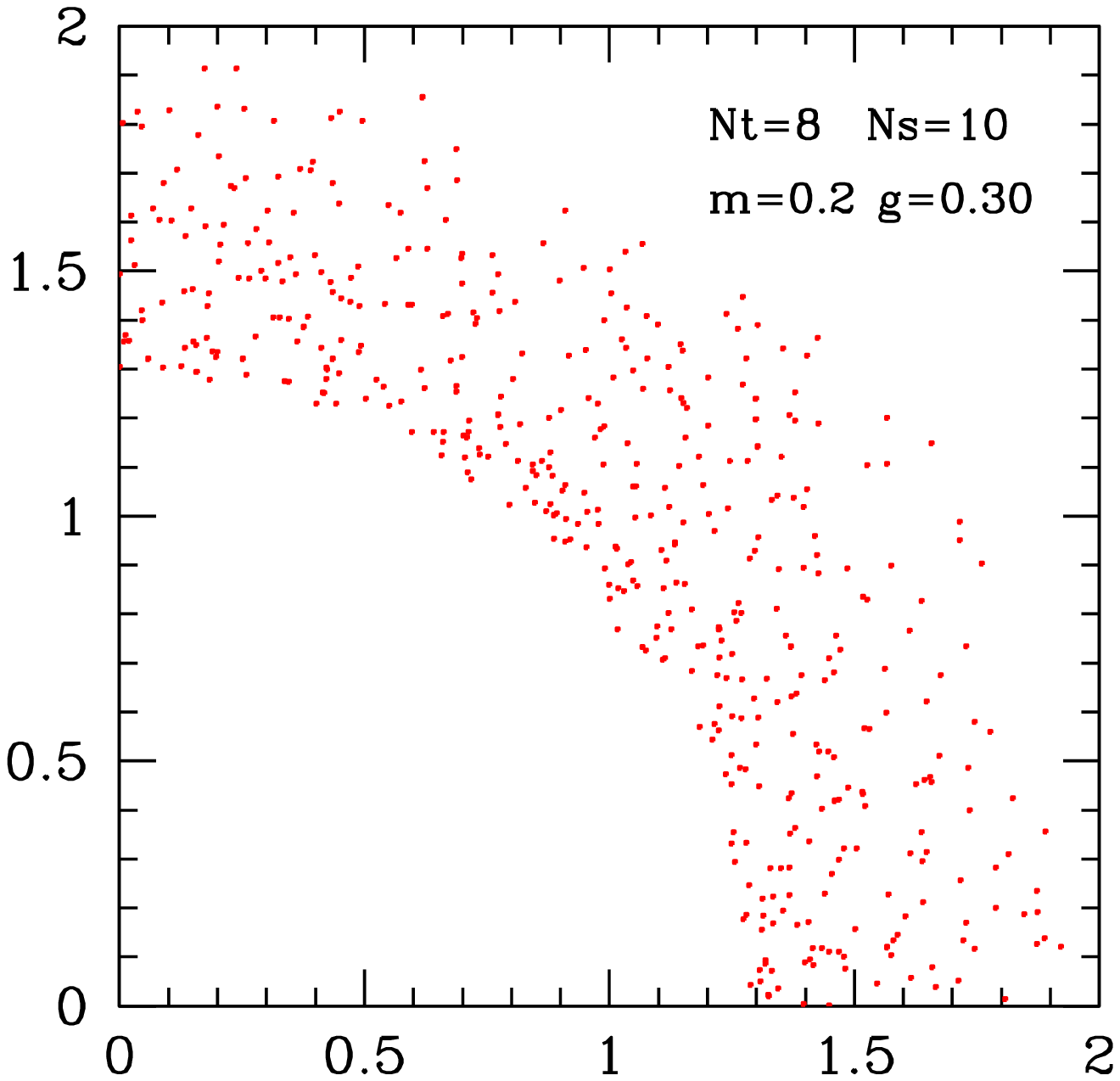,width=8cm,angle=0}
}
\vspace{-2.0cm}
}
\caption{
Eigenvalues in the fugacity plane for $N_t=8$, $N_s=10$, $g=0.30$ and 
$m=0.0$, $0.1$, and $0.2$ respectively, from left to right.
Only the ones with $| \xi_k | \leq 1$ are plotted.
There is  gap around $1$ which increases with $m$.
\label{evals_m}
}
\end{figure}

\subsection{Partition function zeros}

The zeros of the partition function in exactly solvable models
provide an interesting link between finite size and continuum
properties. In the present case one can easily calculate them by expanding 
(\ref{pfdoubleint}) in powers of $e^{\mu}$ or $m$.

It is a well known phenomenon which we have also 
discussed in some technical detail in \cite{RMglasgow}, that
the zeros trace out the analytic continuation of the phase boundaries 
into the complex plane of the respective model parameter.

\begin{figure}
\centerline{
\vbox{
\hbox{
\psfig{file=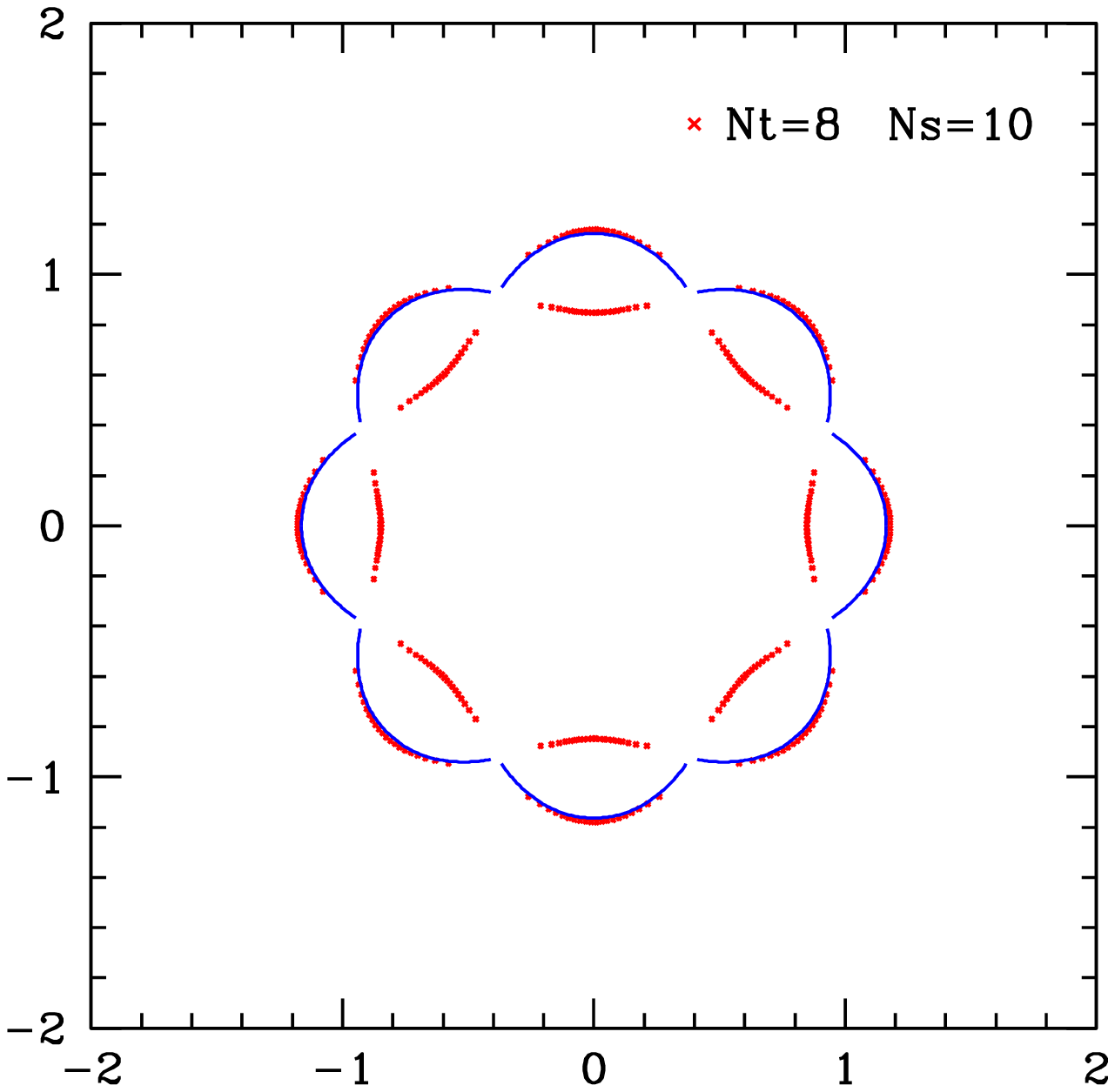,width=7cm,angle=0}
\hspace{-1.0cm}
\psfig{file=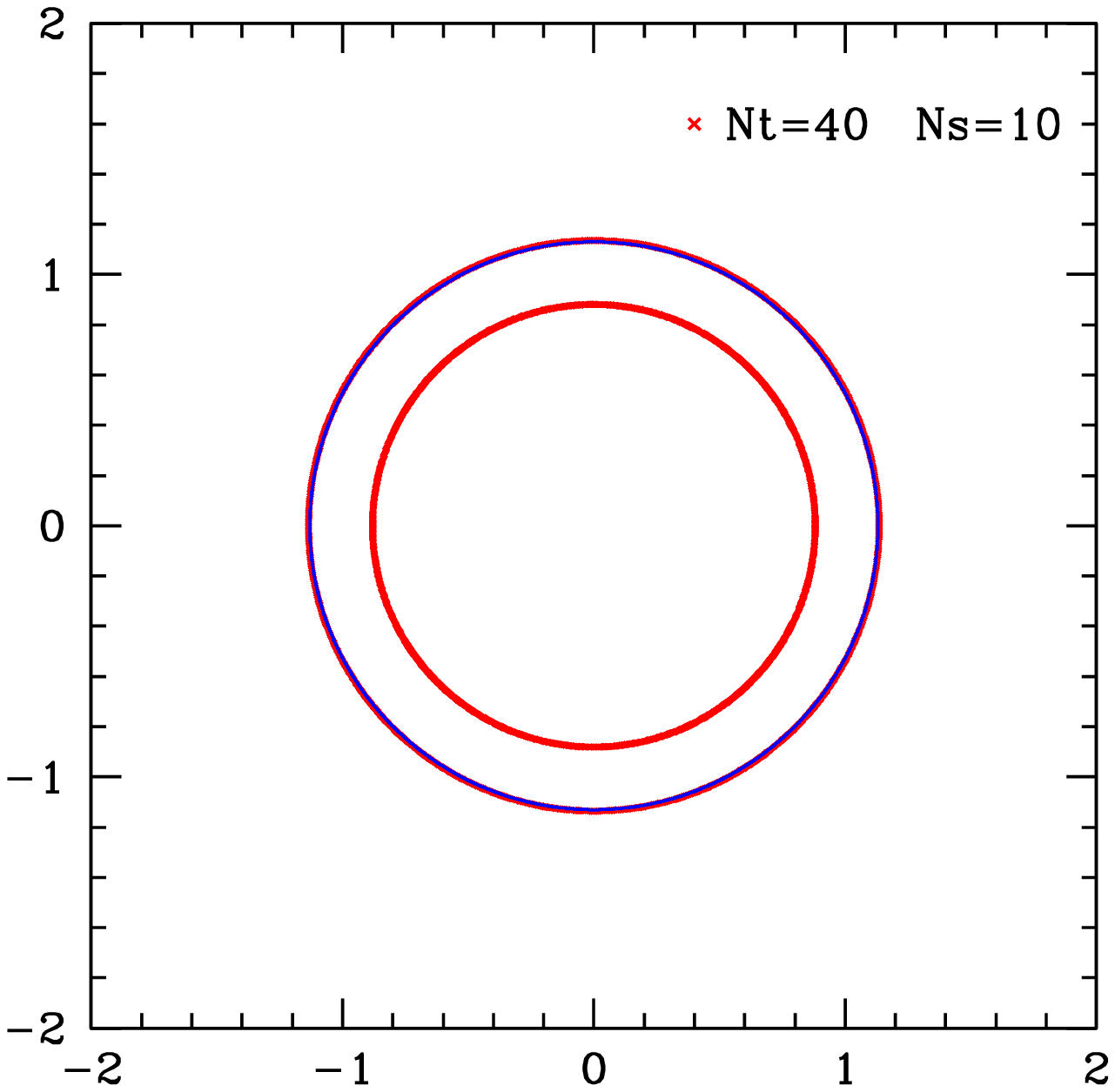,width=7cm,angle=0}
}
\vspace{-2.0cm}
\hbox{
\psfig{file=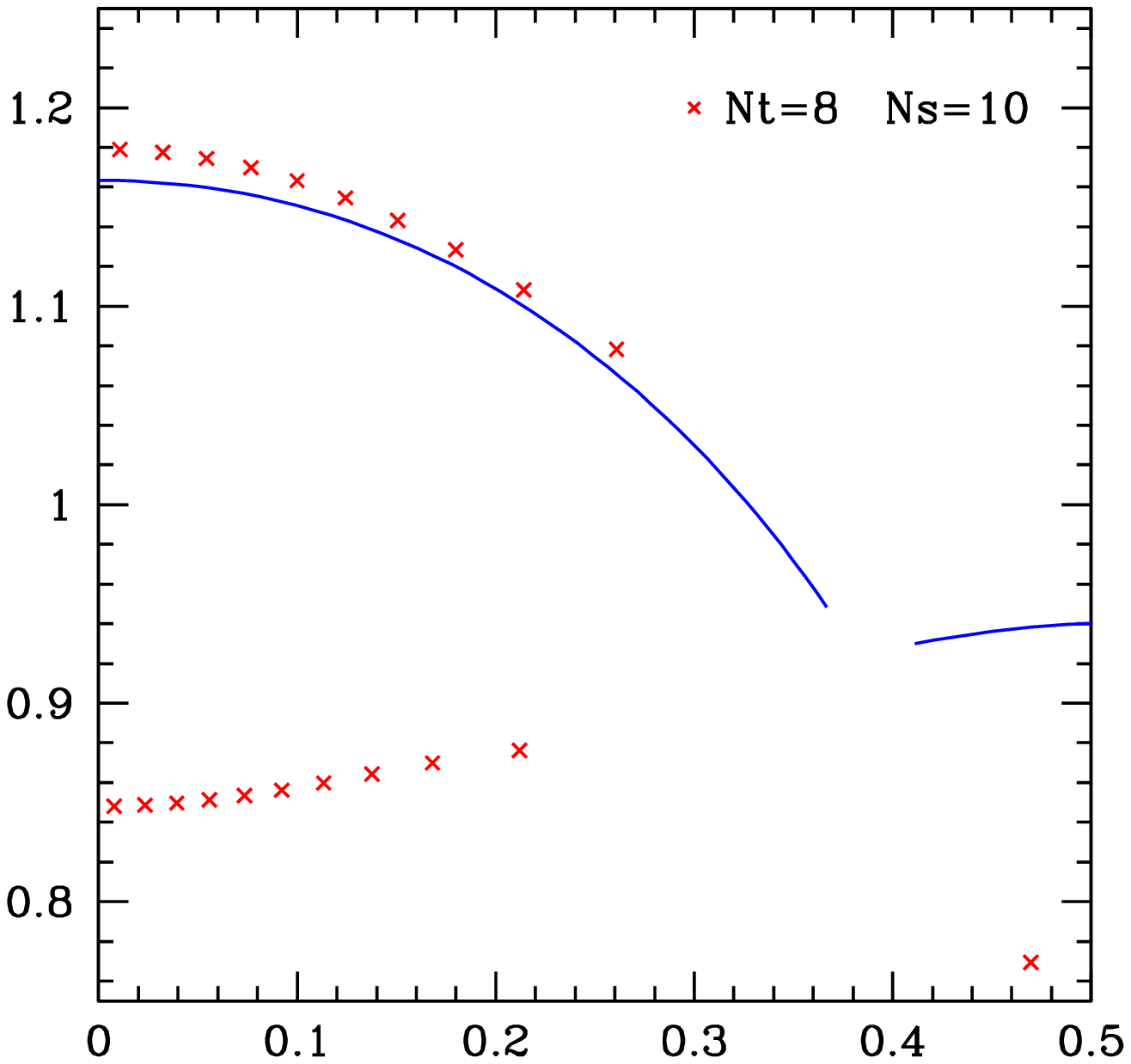,width=7cm,angle=0}
\hspace{-1.0cm}
\psfig{file=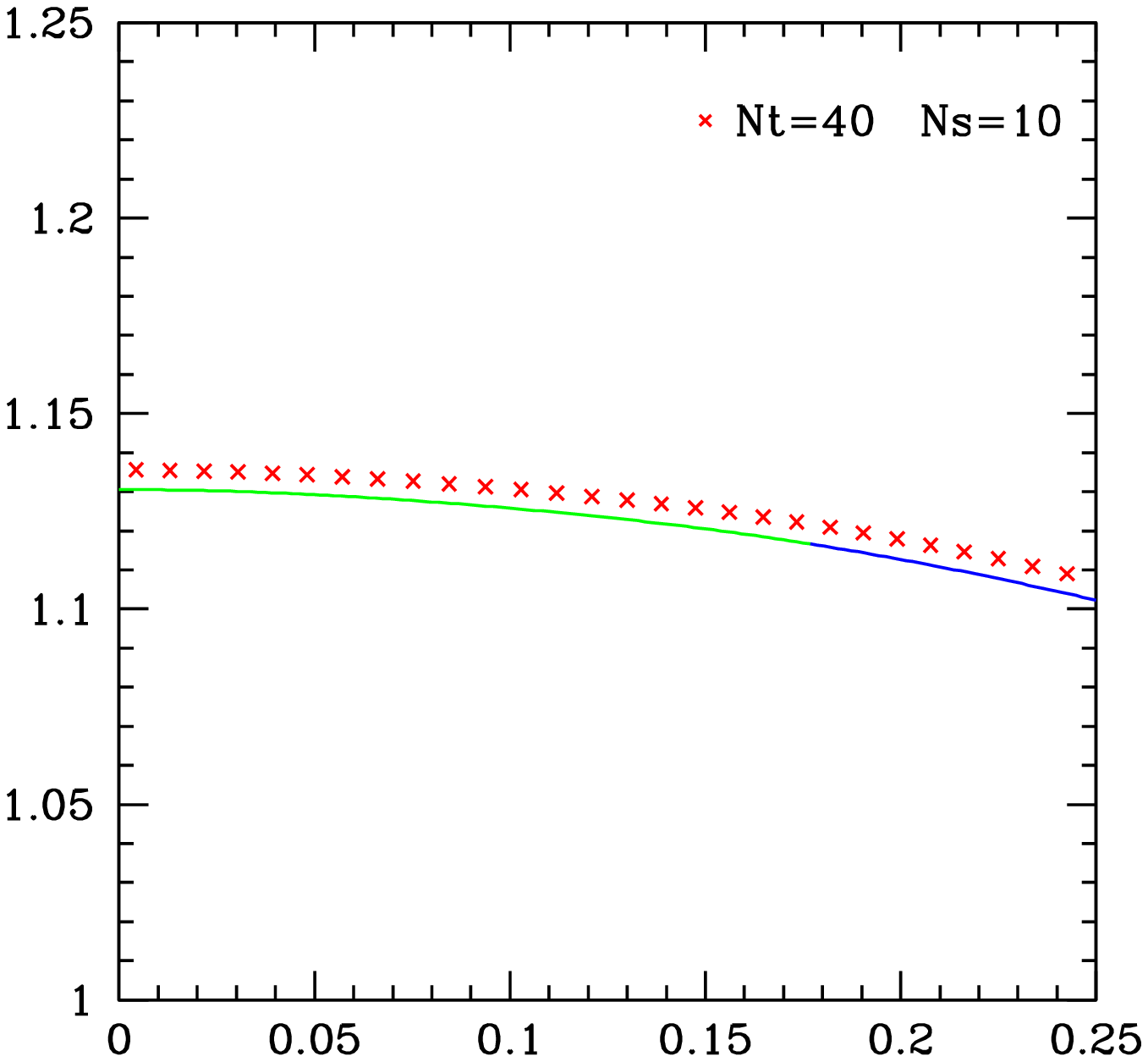,width=7cm,angle=0}
}
\vspace{-2.0cm}
}
}
\caption{
Partition function zeros and the analytical curve that separates
the two saddle points in the complex $\mu$ plane.
The curve within the unit circle is just the 'mirror image' of
the one outside, since everything is symmetric under $\mu \rightarrow 1/\mu$.
In the limits of large $N_t$, the curves become circles at $| \mu | = \mu_c $
and $1/\mu_c$ . The partition function zeros trace out nicely the saddle
point result, even for a moderate $N_s= 10$.
\label{zeros}
}
\end{figure}

This is illustrated in Figure \ref{zeros}, where we show the partition
function zeros in the fugacity plane at zero mass, for the same $N_s=10$
and two different $N_t$ values, $8$ and $40$. The smaller value corresponds
to Glasgow simulations \cite{glasgow}, where they used eight timeslices.
Notice that the zeros are significantly away from a circle, which would be
the continuum result. In the same figures we plot the analytical line corresponding to the transition in the complex fugacity plane, for the same $N_t$ values, but using the saddle point approximation in $N_s$.

\begin{figure} 
\centerline{
\vbox{
\psfig{file=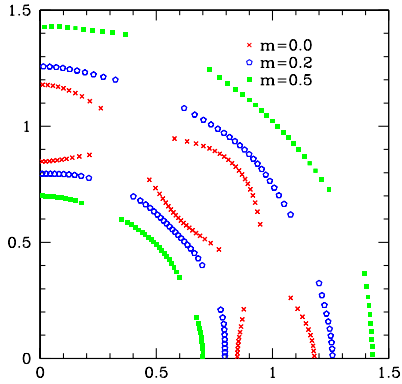,width=8cm,angle=0}
\vspace{-2.5cm}}}
\caption{
Partition function zeros in the complex fugacity plane,
for $N_t=8$ and $N_s=10$, for various mass values.
\label{zerosmu_vsm}
}
\end{figure}

\begin{figure}
\centerline{
\vbox{
\psfig{file=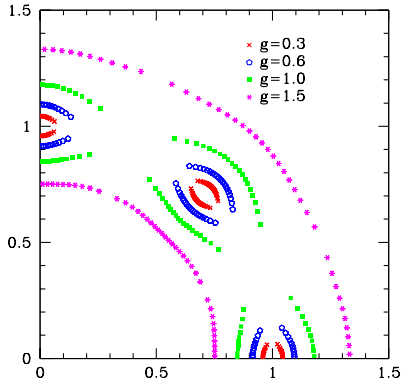,width=8cm,angle=0}
\vspace{-2.5cm}}}
\caption{
Partition function zeros in the complex fugacity plane,
for $N_t=8$ and $N_s=10$, for various values of the size parameter $g$.
\label{zerosmu_vsg}
}
\end{figure}

In Figures \ref{zerosmu_vsm} and \ref{zerosmu_vsg} we plot partition function
zeros for $N_t=8$ and various values of the quark mass $m$ and the size 
parameter $g$. As the mass is increased, the two lines of zeros move away
from the unit circle. For small values of the size parameter $g$, the zeros 
sit in loops around the $N_t$ order complex roots of unity.

\section{Conclusions}

We have constructed a random matrix model for the chiral phase transition
at finite chemical potential with a $\mu$ dependence more similar to
that in the lattice implementation of the problem. Our approach may be interpreted
as ensuring time translation invariance in the large $N_t$ limit. 
We analyzed the model in the saddle point approximation with respect to
the number $N_s$ of points per timeslice (or number of instanton modes).
Similarly to the usual RMM, our model features a first-order phase transition 
driven by the chemical potential $\mu$. Both the particle number $\langle n 
\rangle$ and the chiral condensate have a finite jump at this transition.
The finite jump, hence the first order character, survive when the quark
mass is increased.

We calculated the eigenvalues of the Dirac operator in both
the complex quark mass plane and the complex plane of the fugacity, $e^\mu$.
The distributions of the eigenvalues show a strong qualitative similarity to those
encountered in lattice simulations. 
We also calculated the zeros of the partition function in the fugacity plane
for various parameter values and found that their location is well approximated by 
the phase transition curve obtained from the saddle point analysis of the effective
partition function.

The most remarkable property of the model is the $\mu$-independence of the
partition function in the low density phase, when the number of timeslices $N_t$ is large.
As a consequence, the number density is identically zero in this phase.
This property is different from what is seen in the traditional RMM 
\cite{tricritical,mupaper}, where
there was a negative number density in the low density phase, which coincided
with a suppression of the partition function. Our previous conclusion on the
mechanism that made the Glasgow method exponentially slow points precisely towards 
this suppression as the ultimate source of trouble \cite{RMglasgow}. 

For small $N_t$ values the suppression of the partition function is significant 
in this model, too.
However, it disappears quickly as $N_t$ increases,
together with the $\mu$ dependence in the broken phase, 
as one would expect in continuum QCD.
The Glasgow lattice simulations which have been reported \cite{glasgow} employed
a small number of timeslices, corresponding to $N_t=8$, $N_t'=4$ in our model.
For this value, we find a significant suppression of the partition function. 
The suppression is further enhanced by the number of spatial points, corresponding
to our $N_s$, which is of order $N_t^3$ in a traditional lattice simulation. 
This leads to the exciting possibility that lattice (Glasgow) simulations of 
QCD at finite $\mu$ 
might be plagued by the same problem of a spuriously suppressed partition function,
as one finds in the RMM.
The suppression is an artifact of time discretization, and could 
in principle be cured
by using a large number of timeslices and a small spatial lattice.

The main argument against excessive enthusiasm is the fact that in the present
model, similarly to the usual RMM, we have a major cancellation due to averaging.
Averaging over the phases of the determinant achieves a 
cancellation in the broken or low density phase, namely from the magnitude of the typical
determinant, of the order of $\exp ( 2 \mu N )$, a large number, 
to that of the same determinant at $\mu=0$, which is of the order $1$. 
This cancellation appears then to be similar to the one seen in the usual RMM.
However, in that case the determinant at $\mu \ne 0$ was not significantly larger than $1$, 
and the cancellation due to phase averaging led to a small number, $\exp ( - \mu^2 N )$.
That cancellation coincides with the suppression of the partition function at $\mu\ne 0$
with respect to its value at $\mu=0$.
As we have learned from the analysis of the Glasgow method applied to that model, 
a partition function which is suppressed for some parameter values is very hard 
to calculate because of the huge precision required.

One should make a distinction between the cancellation brought about by
phase averaging, i.e., from $< |det(\mu)| >$ to $<det(\mu)>$, and the suppression of the
partition function, i.e., from $<det(0)>$ to $<det(\mu)>$. 
A partition function like the present one, which is the result
of large cancellations during averaging, but does not have significant suppression, in not a priori incalculable from simulations, and might be amenable to a calculable form
using an appropriate resummation \cite{cluster}.

Our previous study of Glasgow averaging \cite{RMglasgow}
in the usual RMM indicates the suppression of the partition function, $Z(\mu)/Z(0) << 1$
 as the source of trouble. The pattern of false zeros was not
driven by the competition between the true partition function and the Stephanov phase 
(defined by the absolute value of the determinant), but rather between the former and a phase
resulting from spoiling the cancellation between various terms of the polynomial expansion
of the true partition function. 
It is probably too early at this point to encourage a new set of lattice Glasgow simulations.
However, a reexamination of the Glasgow method in the context of this model
might lead to encouraging results. 

\bigskip

Our model is similar to those arising from strong-coupling analysis of lattice
QCD at finite $\mu$ \cite{strongcoupling}, 
but it has the simplicity of random matrix models.
There is a straightforward connection with RMM models with many
Matsubara frequencies \cite{manyMats}. This can be seen by diagonalizing the hopping matrix
in the original form of the partition function.
Extracting a  $T-\mu$ phase diagram from this model is also possible. 
Introducing a dimensionful lattice spacing $a$
leads to identifying the quantity $a N_t$ with the inverse temperature.
However, a more careful analysis of the implementation of temperature in 
this model should precede that.

In summary, we studied a simple model which shows a clear connection between 
time translation invariance and the physical requirement of having a zero
baryon number density in the low density phase of QCD. The careful 
implementation of time dependence  seems to have the potential to 
cure one of the artifacts that make traditional lattice simulations 
at finite density virtually impossible. 

It would be interesting to see how the introduction of an explicit temperature 
dependence in this model would change the $\mu-T$ phase diagram that emerges from the usual
random matrix model. Also, it would be very interesting to see how the exciting
results obtained in the RMM for color superconductivity \cite{colorRM} would be influenced by
our method of introducing the chemical potential.

\bigskip

This work has been supported in part by a grant from the U.S. National
Science Foundation. Thanks are due to James Osborn, Dirk Rischke, Dominique
Toublan, and Jac Verbaarschot for many useful discussions.
Ralph Amado, Dirk Rischke and Jac Verbaarschot are thanked for a critical reading of the manuscript.
I am grateful to the Nuclear Theory Group at BNL for their hospitality during
part of the time the paper was written.

\appendix

\section{Derivation of $\det \lp Q_N ( a , x ) \rp$}

We wish to study matrices of the form
\ber
{\cal Q}_L(a,x)~=~\lp \bacccc a    &    x   & \cdots & x^{-1} \\ 
                     x^{-1} & \ddots & \ddots & \vdots \\
                     \vdots & \ddots & \ddots &  x \\
                        x   & \cdots & x^{-1} &  a \ea \rp~~=~~
a \otimes \unit_N + x \otimes \step_N + x^{-1} \otimes \step_N^T.
\eer
which occur in the dual representation of our partition function.
This result can also be found in \cite{gibbs1}. We derive it here for completeness.
The size of the matrix $N$ is an arbitrary positive integer. There are no
special restrictions on $a$ and $x$. This determinant is 
We will calculate the determinant by induction. First, we define an
auxiliary matrix,
\ber
\tilde{\cal Q}_N(a,x)~=~\lp \bacccc a    &    x   & \cdots & 0 \\ 
                     x^{-1} & \ddots & \ddots & \vdots \\
                     \vdots & \ddots & \ddots &  x \\
                        0   & \cdots & x^{-1} &  a \ea \rp~~.
\eer
The respective determinants $Q_N$ and $\tilde{Q}_N$ verify
\ber
\label{QtQ}
Q_N ~=~ a \tQ_{N-1} - 2 \tQ_{N-2} - (-)^N \lp x^N + \frac{1}{x^N} \rp~~,
\eer
which is obtained as follows. First expand $Q_N$ by its first row,
which leads to three sub-determinants, one of them $\tQ_{N-1}$. 
Expand again the remaining two by their first column, which leads to
two matrices $\tQ_{N-2}$ and triangular matrices, which give the $x^{\pm N}$
terms. The same trick applied to $\tQ_N$ yields a recursion relation,
\ber
\label{tQrec}
\tQ_N ~=~ a \tQ_{N-1} - \tQ{_N-2}~~.
\eer
We have $\tQ_2=a^2 -1 $ and $\tQ_3 = a^3 - 2 a$, which leads to $\tQ_0=1$
and $\tQ_1=a$.
The recursion formula (\ref{tQrec}) can be solved explicitly. First, notice
that the recursion is homogeneous so if two distinct series $\lb q_k \rb$ and
$\lb q'_k \rb$ verify the recursion relation, so will any linear combination
of the two series. Of course, there may be at most two linearly independent
solutions, since the first two terms of a given solution completely define 
the whole solution. We seek the solution in the form of a power series,
$q_k=A \xi^k$. The base $\xi$ must then verify the characteristic equation
\ber
\label{chareq}
\xi^2-a \xi + 1 =0~~.
\eer
Consider the two roots of (\ref{chareq}),
$\lambda_{12}(a)=\frac12 \lp a \pm \sqrt{a^2 - 4} \rp$. They are
each others' inverse, $\lambda_1(a)=\frac{1}{\lambda_2(a)}$, 
and their sum is $a$. We can then write 
$\tQ_k = \alpha_1 \lambda_1^k + \alpha_2 \lambda_2^k$, 
where the coefficient-a $\alpha_{12}$ are determined from the first 
two terms in the series, as
\ber
\alpha_1~=~\frac{\lambda_1}{\lambda_1 - \lambda_2}~~;~~
\alpha_2~=~\frac{-\lambda_2}{\lambda_1 - \lambda_2}~~.
\eer
The result $\tQ_N = \frac{\lambda_1^{N+1} - \lambda_2^{N+1}}{\lambda_1-
\lambda_2}$ finally leads to the following expression for the original
determinant:
\ber
\label{Qexpl}
Q_N~=~\lambda^N + \lambda^{-N} - (-)^N \lp x^N + x^{-N} \rp~~.
\eer
where for simplicity we defined
$\lambda \equiv \lambda_1 = \frac12 \lp a \pm \sqrt{a^2 - 4} \rp 
= \frac{1}{\lambda_2}$.


\end{document}